\documentclass[aps,prl,reprint,showpacs,superscriptaddress]{revtex4-1}

\usepackage{graphicx}
\usepackage{subfigure}
\usepackage{color}
\usepackage{amsmath}
\usepackage{amssymb}
\usepackage{epstopdf}
\usepackage[linktocpage,colorlinks=true,linkcolor=blue,citecolor=blue,breaklinks=true]{hyperref}
\usepackage{verbatim}
\usepackage{breakcites}
\usepackage[version=3]{mhchem}

\newcommand{\be}{\begin{equation}}
\newcommand{\ee}{\end{equation}}

\begin{document}

\preprint{}

\title{Theory of the sea ice thickness distribution}

\author{Srikanth Toppaladoddi}
\affiliation{Yale University, New Haven, USA}
\affiliation{Mathematical Institute, University of Oxford, Oxford, UK}

\author{J. S. Wettlaufer}
\affiliation{Yale University, New Haven, USA}
\affiliation{Mathematical Institute, University of Oxford, Oxford, UK}
\affiliation{Nordita, Royal Institute of Technology and Stockholm University, SE-10691 Stockholm, Sweden}

\email[]{john.wettlaufer@yale.edu}

\date{\today}

\begin{abstract}
We use concepts from statistical physics to transform the original evolution equation for the sea ice thickness distribution $g(h)$ due to \citet[][]{Thorndike:1975} into a Fokker-Planck like conservation law.   The steady solution is $g(h) = {\cal N}(q) h^q \mathrm{e}^{-~ h/H}$, where $q$ and $H$ are expressible in terms of moments over the transition probabilities between thickness categories. The solution exhibits the functional form used in observational fits and shows that for $h \ll 1$, $g(h)$ is controlled by both thermodynamics and mechanics, whereas for $h \gg 1$ only mechanics controls $g(h)$.  Finally, we derive the underlying Langevin equation governing the dynamics of the ice thickness $h$, from which we predict the observed $g(h)$.  
The genericity of our approach provides a framework for studying the geophysical scale structure of the ice pack using methods of broad relevance in statistical mechanics.

\end{abstract}


\maketitle

\vspace{-0.7 in}

The Earth's climate system is a complex nonlinear dynamical system \cite{Dijkstra}.  Three main research approaches in climate science are common:  (a) Observation of the past and present state of the system and extrapolation to the future.  (b) Numerical simulations using Global Circulation Models, which treat the system with the deterministic approach of weather forecasting by modeling the processes on a coarse-grained scale. (c) Constructing a low-order description of the system/subsystem of the climate, in the vein of theoretical physics.   There are substantial cultural and technical differences between these approaches.  All have value and all have limitations.  Evidently, the ready availability of computing power has made approach (c) less favorable.  Here we show that one of the key variables in polar climate, the sea ice thickness distribution $g(h)$, can be fruitfully examined quantitatively with a core set of tools in statistical physics.

Although there are hi-fidelity measurements of the area of the ice cover during the satellite era \cite{OneWatt}, the key quantity reflecting the climatological state of the sea ice cover is its volume, making $g(h)$ the central state variable of the system.  The thickness distribution underlies and reflects ice melting or freezing due to the thermodynamic 
forcing of the ocean and atmosphere, and mechanical deformation; rafting, ridging and the formation of open water \citep{Thorndike:1975}. 
Nonetheless, although the theory of the ice thickness distribution has been with us for forty years we still seek a basic understanding of its components in order to test its predictions \citep{OneWatt}.  

The theory of \citet{Thorndike:1975} is described by a continuous deterministic partial differential equation that contains the principal physical processes mentioned above and is given by
\be
\frac{\partial g}{\partial t} = - \nabla \cdot ({\bf u} g) - \frac{\partial}{\partial h} \left(f g\right) + \psi,
\label{eqn:thorndike}
\ee
where ${\bf u}$ is the velocity of the ice pack and $f$ is its growth/melt rate.   The principal reason for the difficulty in testing the theory arises from the so-called redistribution function, $\psi$, intended to capture mechanical processes. Although from observational, theoretical and numerical perspectives we  have gained a quantitative explanation for many aspects of the the redistribution function \cite[e.g.,][and refs. therein]{Thorndike:1992, Thorndike:2000}, a closed mathematical analysis of the original theory is still principally limited by this term.  In what follows, by viewing $\psi$ within the framework of kinetic theory, we show that the original theory can be rewritten as a Fokker-Planck type equation.  In so doing a number of useful advantages arise.  First, we can determine the steady solution analytically.  Second, we provide access to the full range of methods and approaches of non-equilibrium statistical physics.  In this note we describe just two; (1) by comparison with observations we deduce the transport  coefficients in the new evolution equation, which allow  for its full numerical solution in a geophysically relevant setting and (2) we derive the corresponding Langevin equation for the evolution of the ice thickness itself.  
 

A central ansatz in stochastic dynamics is to consider the ``microscopic noise'' underlying a ``macroscopic process'' as decorrelating on a time scale far faster than the macroscopic displacement.  The classical test is Brownian motion, 
wherein the inertial macroscopic displacement of a pollen grain in water evolves slowly relative to the collisional events driving its motion.  This is embodied in the fluctuation-dissipation theorem \cite[e.g.,][]{Reif}. 

Within this framework we recast $\psi$ as follows.   Each of the ``microscopic'' mechanical processes that influence the ice thickness distribution--rafting, ridging and the formation of open water--occur over a time scale that is very rapid relative to the geophysical-scale changes of $g(h)$.  We thus view these processes as the collisions of solvent molecules with a Brownian particle; the individual collisions have a probability of displacing the particle, but their phase space is so enormous that we do not study them individually.  In the same vein, we do not study the individual floe--floe interactions in the ice pack.  Rather, we write $\psi$ as
\be
\psi = \int_{0}^{\infty} \left[g(h+h^\prime) w(h+h^\prime,h^\prime) - g(h) w(h,h^\prime)\right] dh^\prime.
\label{eqn:psi}
\ee
Here, the first and second terms represent the processes by which (i) ice floes of thickness $h+h^\prime$ become ice floes of thickness $h$ and (ii) ice floes of thickness $h$ become ice floes of thickness $h-h^\prime$ respectively, with $w~dh^\prime$ the transition probability per unit time for these events.

Taylor expanding the right hand side of equation (\ref{eqn:psi}) and substituting this is into equation (\ref{eqn:thorndike}), we obtain
\be
\frac{\partial g}{\partial t} = - \nabla \cdot ({\bf u} g) - \frac{\partial}{\partial h} \left(f g\right) + \frac{\partial}{\partial h} (k_1 g) + \frac{\partial^2}{\partial h^2} (k_2 g),
\label{eqn:newg(h)}
\ee
where
\be
k_1 = \int_{0}^{\infty} h^\prime w(h,h^\prime) dh^\prime \mathrm{~and~}
k_2 = \int_{0}^{\infty} \frac{1}{2}{h^\prime}^2 w(h,h^\prime) dh^\prime.
\label{eqn:coeffs}
\ee
Thus, we have transformed the original theory into a Fokker-Planck type of evolution equation. Note that in the absence of ice motion equation (\ref{eqn:newg(h)}) is exactly the Fokker-Planck equation; an advection-diffusion equation for the probability density \cite{Reif, Dijkstra}.  Here, the coefficients (Eqs. \ref{eqn:coeffs}) are the first and second order moments of the transition probability between ice thickness categories.

Choosing $L$ as the horizontal scale, $H_{eq}$ as the vertical scale and $U_0$ as the velocity scale, we find
three time scales in the problem; (1) the thermal diffusion time scale, $t_D = H_{eq}^2/\kappa$, with $\kappa$ the thermal diffusivity of ice; (2) the time scale associated with the horizontal motion of ice floes, $t_m = L/U_0$; (3) the relaxation time scale of the ice floes when they are involved in collisions, $t_R \sim 1/\dot \gamma$, where $\dot \gamma$ is the collisional strain rate. These time scales are such that, $t_m \approx t_R$ and $\tau \equiv t_R/t_D \ll 1$. Hence, we have 
$f_0 = H_{eq}/t_D$, $\widetilde{k_{1}} = H_{eq}/t_R$, and $\widetilde{k_{2}} = H_{eq}^2/t_R$ as the scales for the remaining terms.  Maintaining the prescaled notation, the dimensionless equation can be written in one spatial dimension as
\be
\frac{\partial g}{\partial t} = -  \frac{\partial}{\partial x} (ug) - \tau \frac{\partial}{\partial h} \left(f g\right) + \frac{\partial }{\partial h} \left(k_1 g\right) + \frac{\partial ^2}{\partial h^2} \left(k_2 g\right).
\label{eqn:1D}
\ee

Now we obtain the steady solely $h$-dependent solution of equation (\ref{eqn:1D}) with boundary conditions $g(0) = g(\infty)=0$.  The growth rate $f$ in the original theory of \citet[][]{Thorndike:1975} was determined numerically from the climatologically forced Stefan problem for the ice thickness.  If $\Delta T$ is the temperature difference over a solid layer of thickness $h$, we take a standard analytical solution for its diffusive growth into an isothermal liquid \cite[e.g.,][]{Worster:2000}.  Here, one balances heat conduction through the layer ($\propto {\Delta T}/h$) against latent heat production at the interface ($\propto dh/dt \equiv f$) giving 
\be
f = \frac{1}{S} \left(\frac{1}{h}\right),
\ee
with $S = L/c_p \Delta T$ the Stefan number, in which $L$ is the latent heat of fusion and $c_p$ is the specific heat at constant pressure.   Ignoring the advection term this leads to
\be
\frac{\partial g}{\partial t} = -\frac{\partial}{\partial h} \left[\left(\frac{\epsilon}{h} - k_1 \right)g\right] + \frac{\partial^2}{\partial h^2} \left(k_2 g\right)
\label{eqn2}
\ee
where $\epsilon \equiv \tau/ S \ll 1$ because $S \gg 1$ and $\tau \ll 1$.  Because the small parameter multiplies regular singularities, which become $O(1)$ when $h = O(\epsilon)$, we keep all terms in equation (\ref{eqn2}), to which we seek the stationary solution and rewrite as
\be
\frac{d^2g}{dh^2} + \frac{d}{dh} \left[\left(\frac{1}{H} - \frac{q}{h}\right) g \right] = 0,
\ee
where $H = k_2/k_1$ and $q = \epsilon/k_2$. The first integral is
\be
\frac{d g}{dh} + \left(\frac{1}{H} - \frac{q}{h}\right) g = B,
\label{eqn3}
\ee
where $B$ is the integration constant. We solve equation (\ref{eqn3}) using an integrating factor $e^{h/H-q\ell n(h)}$, which requires $B = 0$ to satisfy $g(0) = g(\infty)=0$,
and find that
\be
g(h) = {\cal N}(q) h^q e^{-h/H}. 
\label{eqn:soln1}
\ee
The prefactor is determined by the normalization condition $\int_0^{\infty} g(h) dh = 1$, and is ${\cal N}(q) = \left[H^{1+q} \Gamma(1+q)\right]^{-1}$, with $\Gamma(x)$ the Euler gamma function.  Hence ${\cal N}(q)$ is unique and single valued for ${\mathbb{R}}(q) > -1$ and $\mathbb{R}(H) > 0$.  Note that $q$ and $H$ have an independent interpretation within the framework of the theory {\em and} are the sole fitting parameters. Clearly, for $h \ll 1$, $g(h)$ is controlled by both thermodynamics and mechanics, whereas for $h \gg 1$,  $g(h)$ is controlled solely by mechanical interactions.

A Fokker-Planck equation describes the evolution of the probability density of a random process, but to study the random process itself ($h$ in our case) we study the Langevin equation corresponding to equation (\ref{eqn2}), which we write as
\be
\frac{dh}{dt} =  \left(\frac{\epsilon}{h} - k_1 \right) + \sqrt{2k_2}~\xi(t),
\label{eqn:langevin1}
\ee
where $\left(\frac{\epsilon}{h} - k_1 \right)$ and $\sqrt{2k_2}$ are the drift and diffusion terms respectively, and $\xi(t)$ is Gaussian white noise  \cite[e.g.,][]{stratonovich1969}.
Clearly, our assumption of $k_2$ being a constant in the determination of the solution for $g(h)$ in equation (\ref{eqn:soln1}), translates into 
additive noise in the corresponding Langevin equation shown in (\ref{eqn:langevin1}).

\begin{figure}
\centering
\includegraphics[trim = 0 0 0 0, clip, width = 1\linewidth]{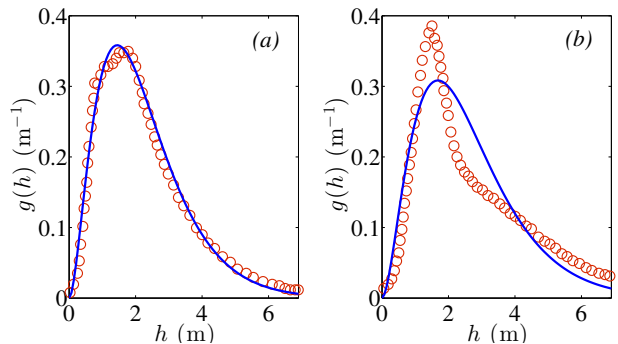}
\vspace{-0.25 in}
\caption{Comparison of our theory with satellite measurements for February-March (F-M) of  (a) 2008 and (b) 2004. Circles are the distribution functions from ICESat \cite{kwok2009} and lines are the fits using equation (\ref{eqn:soln1}). In (a), $q=1.849$ and $H=0.783$ m, and in (b), $q=1.848$ and $H=0.910$ m.}
\vspace{-0.25 in}
\label{fig:icesat}
\end{figure}

We now compare our theory with the thickness distributions obtained during the ICESat mission \cite{kwok2009}. Figure \ref{fig:icesat} shows fits of our solution (\ref{eqn:soln1}) to the distribution functions for the period February-March (F-M) for (a) 2008 and (b) 2004, which we have chosen to demonstrate both the typical fit (2008) and the worst fit (2004) of our solution to the observations \cite[c.f., Fig. 6 of][]{kwok2009}.   The key reasons for deviations are (1) the observations span the ice cover, and yet near landmasses, and depending on wind direction, it is possible that $k_1$ and $k_2$ will differ locally, (2) we neglected the advection term  when determining the solution, (3) the form of $f$ used in equation (\ref{eqn:newg(h)}) is the solution of the ideal Stefan problem for {\em growth} only, and (4) we are comparing the steady solution to the data.  Incorporating these and related issues are key aspects of a thorough numerical study of equation \ref{eqn:1D}, which are part of a longer treatment.

\begin{figure}
\centering
\includegraphics[trim = 0 0 0 0, clip, width = 1\linewidth]{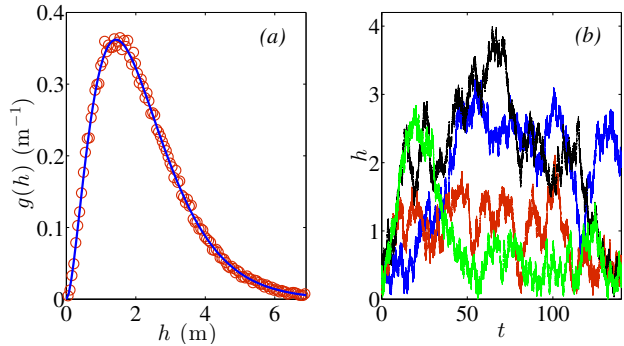}
\vspace{-0.25 in}
\caption{Solution to the Langevin equation (\ref{eqn:langevin1}) for F-M 2008. Here, $\epsilon = 0.046$, $k_1 = 0.048$ and $k_2 = 0.025$. The ensemble size used is $N_{en}=10^5$ and the time step for the Euler-Maruyama scheme \cite{higham2001} is $\Delta t = 10^{-5}$. The total non-dimensional integration time is $T=140$. Figures (a) and (b) show the thickness distribution and four realizations from the ensemble respectively. In (a), the circles represent $g(h)$ from the Langevin equation and the solid curve is the fit using equation (\ref{eqn:soln1}), which gives $q = 1.838$ and $H = 0.777$ m.}
\label{fig:langevin}
\vspace{-0.25 in}
\end{figure}

Now from the values of $q$ and $H$ we obtain $k_1$ and $k_2$ and use them in equation (\ref{eqn:langevin1}) to evolve $h$ itself. The solution to the Langevin equation corresponding to figure \ref{fig:icesat}(a) is shown in figure \ref{fig:langevin}. Invoking ergodicity, clearly equation (\ref{eqn:langevin1}) qualitatively reproduces the observed $g(h)$ and thus acts as an ideal and simple model to study the thickness distribution.  

We have transformed the original evolution equation for the sea ice thickness distribution, $g(h)$, due to \citet[][]{Thorndike:1975}, to a Fokker-Planck like equation by recasting the redistribution function, $\psi$, using the analogy with the theory Brownian motion. The idea is that the mechanical processes embodied in $\psi$ (rafting, ridging and the formation of open water) are thought of like the collisions of solvent molecules with a Brownian particle--the individual events that change the ice thickness occur on time and space scales that are short relative to the geophysical-scale changes of $g(h)$. Thus, we do not treat the individual floe--floe interactions in the ice pack, but rather only the moments of the transition probabilities for these events.  That the integrals describing these moments rapidly converge to take constant values is borne out by comparison with observations; the stationary solution (equation \ref{eqn:soln1}) of the new evolution equation captures the basin-scale satellite measurements of the distribution.
Finally, the corresponding Langevin equation (\ref{eqn:langevin1}) is evolved with observationally constrained parameters to study the evolution of $h$ itself.  The associated agreement of the $g(h)$ obtained from this approach with the observations is consistent with an ergodic thickness field.  The simplicity of the approach and its immediate connection with the edifice of non-equilibrium statistical mechanics make it appealing for a wide range of reasons, from context for comparison with observations to simplification of models.

\begin{acknowledgements}
ST acknowledges a NASA Graduate Research Fellowship.  JSW acknowledges the Swedish Research Council, a Royal Society Wolfson Research Merit Award and NASA Grant NNH13ZDA001N-CRYO for support.  This work was completed at the 2015 Geophysical Fluid Dynamics Summer Study Program 
``Stochastic Processes in Atmospheric \& Oceanic Dynamics'' 
at the Woods Hole Oceanographic Institution, which is supported by the National Science Foundation and the Office of Naval Research.  We thank the staff for comments and criticisms. 
\end{acknowledgements}
\vspace{-0.25 in}


%

\end{document}